# Capture the growth kinetics of CVD growth of two-dimensional MoS$_2$

*Dancheng Zhu[1], Haibo Shu[2], Feng Jiang[1], Danhui Lv[1], Vijay Asokan[1], Omar Omar[3], Jun Yuan[3,1], Ze Zhang[1] and Chuanhong Jin[1*]*

[1]State Key Laboratory of Silicon Materials, School of Materials Science and Engineering, Zhejiang University, Hangzhou, Zhejiang 310027, China

[2]College of Optical and Electronic Technology, China Jiliang University, Hangzhou 310018, China

[3]Department of Physics, University of York, Heslington, York, YO10 5DD, United Kingdom

Understanding the microscopic mechanism of chemical vapor deposition (CVD) growth of two-dimensional molybdenum disulfide (2D MoS$_2$) is a fundamental issue towards the function-oriented controlled growth. In this work, we report results on revealing the growth kinetics of 2D MoS$_2$ via capturing the nucleation seed, evolution morphology, edge structure and terminations at the atomic scale during CVD growth using the transmission electron microscopy (TEM) and scanning transmission electron microscopy (STEM) studies. The direct growth of few- and mono-layer MoS$_2$ onto graphene based TEM grids allow us to perform the subsequent TEM characterization without any solution-based transfer. Two forms of seeding centers are observed during characterizations: (i) Mo-oxysulfide ($MoO_xS_{2-y}$) nanoparticles either in multi-shelled fullerene-like structures or in compact nanocrystals for the growth of fewer-layer $MoS_2$; (ii) Mo-S atomic clusters in case of monolayer $MoS_2$. In particular, for the monolayer case, at the early stage growth, the morphology appears in irregular polygon shape comprised with two primary edge terminations: S-Mo Klein edge and Mo zigzag edge, approximately in equal numbers, while as the growth proceeds, the morphology further evolves into near-triangle shape in which Mo zigzag edge predominates. Results from density-functional theory calculations are also consistent with the inferred growth kinetics, and thus supportive to the growth mechanism we proposed. In general, the growth mechanisms found here should also be applicable in other 2D materials, such as $MoSe_2$, $WS_2$ and $WSe_2$ etc.

**Introduction**

Two-dimensional molybdenum disulfide (2D $MoS_2$), a representative member of the rediscovered transition metal dichalcogenide (TMDC) family, is holding promising interests owing to their excellent performance in electronic, optoelectronic and catalytic applications.[1-7] However, there are few challenging tasks still remain in reality towards the potential applications, for instance, to modulate the carrier type and concentration i.e., to achieve equivalent good performance for both n- and p-type in electronic devices, to precisely control the structure/morphology, to grow wafer-size sample and in high quality. Out of these, the controlled growth seems to be a key issue. So far, chemical vapor deposition (CVD), as one of the most successful routes, has been widely adopted to grow 2D TMD material from graphene,[8,9] hexagonal boron nitride[10] to 2D TMDC materials,[11-13] while the growth mechanisms, particularly in the CVD growth of 2D TMDC materials, is less understood.

Previous studies have successfully revealed the roles of a number of important parameters determining the CVD growth behaviors of 2D $MoS_2$ via CVD, i.e., hydrogen carrier gas, precursors/promotors and screw dislocations etc.[14-22] Recently Rajan[18] et al. have proposed a generalized mechanistic model to quantitatively explain the shape evolution of $MoS_2$ monolayers observed experimentally. In spite of these successes, our knowledge on the microscopic process during the growth including nucleation and growth kinetics seems to be still limited, partly due to the few practical difficulties, i.e., the challenge to transfer ultra-fine clusters formed at early stage growth, possible loss of intermediate-/by-products (some of them are aqueous soluble). As such, comprehensive studies are highly demanded to reveal the microscopic growth mechanism of 2D $MoS_2$.

In this work, we carried out a CVD experiment and subsequent TEM study to capture the nucleation and subsequent growth kinetics of 2D $MoS_2$. Here few- and mono-layer $MoS_2$ were grown directly on graphene supported TEM grids,[23-26] and then loaded into the microscope chamber for further microscopic characterizations, during which no solution was involved in the transfer process, and thus allowing us to visualize the seeding centers, shape morphology, edge structures and the associated morphology evolved during the growth. We firstly confirmed there exists two form of seeding centers: (1) Mo-oxysulfide ($MoO_xS_{2-y}$, $y \geq x$) nanoparticles either in nested multi-shelled fullerene-like structures or in compact forms of nanocrystals, for few-layer $MoS_2$, and

(2) atomic MoS$_2$ monolayer cluster. For the growth of mono-layer MoS$_2$, it develops from an irregular polygon-shape morphology comprised with two primary forms of edge configurations: S-Mo Klein edge and Mo zigzag edge. As the growth proceeds, it turns to appear in a near-triangular shape predominantly terminated with Mo zigzag edge. Microscopic nucleation and the growth kinetics and the mechanism can be deduced based on our experimental finding, which were further supported by density-function theory (DFT) calculations.

**Results and Discussions**

Figures 1a shows the schematic of our CVD system in which the graphene supported TEM grids (home-made) were placed facing-down to the MoO$_2$-containing boat. Figures 1b and 1c show two low-magnification ADF-STEM images of as-grown MoS$_2$ samples on graphene under two typical growth conditions where the argon (Ar) flow rate (carrier gas) is different: 200sccm (Figure 1b) and 500sccm (Figure 1c). In the former case (low Ar rate), we observed two primary forms of products (Figure 1b): few-layer regions bound with thicker/heavier nanoparticles or nanorods (marked with dotted circles in green) and monolayer MoS$_2$ with a near-triangle shape (see dotted triangle in red). Contrastingly, in the latter case (high Ar rate), as-grown samples contain predominating monolayer MoS$_2$, as shown in Figure 1c.

Further ADF-STEM and energy dispersive X-ray spectra (XEDS) were carried out to probe the atomic structure and the chemical compositions of those nanoparticles- or nanorods-like cores on few-layer MoS$_2$ as shown in Figure 2 and Figure S1a, b. Two primary forms were observed: multi-shell fullerene structure (Figure 2a) or multi-shelled tube- (Figure 2b), with either empty or partially (even completely) filled cores, similar to that of $MoS_2$ inorganic fullerenes;[14,15] compact nanocrystals (Figure 2c) with few-layer samples. Chemical analysis via STEM-XEDS mapping confirm that that it was mostly composed of molybdenum, sulfur and oxygen on the shells, while a higher concentration of oxygen on the compact core. As such we can conclude that those nanoparticles or nanorods bound with few-layer flakes are $MoS_2$ multi-shelled fullerenes (nanotubes) [15] filled with Mo-oxysulfide ($MoO_xS_{2-y}$, $y \geq x$) nanoparticles. During the preparation of this manuscript, we became aware of another work reporting similar fullerene as the seeding materials monolayer $MoS_xSe_{(2-x)}$. [27]

Importantly, another form of nanoparticles rather than the multi-shelled fullerene/tubular structure was experimentally observed which was further proved to be $MoO_xS_{2-y}$ nanocrystals via EDS analysis similar to those cores filled in the fullerenes in Figure 2a. Our experimental findings revealed that these byproducts are either in fullerene structure/tubular, or in the form of compact nanoparticles with the as-formed fewer layer $MoS_2$ (around 100 few-layer samples were checked, 85 of them are in the former form fullerene while the others are in the latter form- compact nanocrystals.). More microscopic studies were carried out to understand the link between them as shown in Figure 3 and Figure S1c in Supplementary Information. Fast Fourier-transform (FFT) show the single crystalline nature of most of few-layer $MoS_2$, in other words, single domain rather than multiple-domain with different crystallography orientations (see Figure S1d). The $MoS_2$ fullerene is filled with two $MoO_xS_{2-y}$ nanocrystals. No obvious linking was found in crystallography orientation relationship between two $MoO_xS_{2-y}$ nanoparticles, $MoS_2$ fullerene and planar few-layer $MoS_2$. In comparison, the $MoO_xS_{2-y}$ compact nanoparticle (red dotted circle) appear to share certain orientation relationship with $(1\text{-}10)_{MoOS}$ // $(1\text{-}10)_{MoS2}$, and identical spacing in $\{1\text{-}10\}$ planes of few-layer $MoS_2$(yellow dotted circle), as shown in Figure 3d and the corresponding FFT patterns. Based on this result, a microscopic process from $MoO_2$ to $MoS_2$ during CVD process can be speculated. As following: $MoO_2$ belongs to tetragonal crystal system with the $\{1\text{-}10\}$ spacing of 0.34nm. During the sulfurization, the intermediate products prefer to reserve their tetragonal structure whiles undergoes a reduction of (1-10) spacing from 0.34nm ($MoO_2$) to 0.27nm ($MoO_xS_{2-y}$, $MoS_2$). As the sulfurization proceeds, the $MoO_xS_{2-y}$ was further transformed into hexagonal few-layer $MoS_2$.

Given the results shown above, we could assign these nanoparticles either fullerene/tubular like structure or compact $MoO_xS_{2-y}$ nanocrystals as the centers for the nucleation and feeding source for the growth. Under this condition, the sublimated molecular clusters of $MoO_2$ are mostly appearing in large sizes, and thus may not be completely sulfurized due to the limited reaction time before their deposition onto graphene substrate. As such, $MoO_xS_{2-y}$ nanoparticles are formed, and then serving as the heterogeneous nucleation sites for the growth of $MoS_2$ in few-layer forms. Such nuclei in large sizes should also facilitate for the nucleation and growth of few-layer $MoS_2$, rather than the monolayer form, either from a few-layer nuclei or through a layer-on-layer growth. As reaction proceeds, the nanoparticles have two different routes to serve as the nucleation sites: (1) The chemical conversion occurs much faster than the diffusion of the sulfur gas into the $MoO_xS_{2-y}$ nanoparticles which will be further sulfurized to form the nested multiple-fullerene nanostructures as shown earlier.[28] As the sulfurization, these as-deposited $MoO_xS_{2-y}$ nanoparticles may also serve as the feeding source for growing $MoS_2$. The fully or partially empty cores observed on those multiple fullerenes may be formed as a result of self-sacrifice as the seeding source. Other mechanisms such as Kirkendall effect [29] during the sulfurization may also lead to the observed empty-core structures. In some cases, the outer shells in high qualities may block the mass transport and thus the sulfurization, leading the central core having a higher concentration of oxygen, as shown in Figure 1b. (2) When the c-axis of these as-deposited $MoO_xS_{2-y}$ nanoparticles is perpendicular to the graphene plane, the particles will serve as the center of epitaxial growth, so the multiple-fullerene nanostructures will not form.

We then turn to the case of monolayer $MoS_2$. Figures 4a to 4l present the ADF-STEM images of the graphene supported as-grown monolayer $MoS_2$ in different sizes and morphology (Note here the $MoS_2$ samples were grown on different regions of graphene support under the same CVD condition, and then grouped in size as a reflection of the structure evolution, but not exactly the growth kinetics of the same $MoS_2$ monolayer sample under the different growth stages). The varying size and morphology of these $MoS_2$ monolayers on different regions of the same graphene substrate may be a result from the local fluctuations responsible for nucleation and growth. In the samples we checked (over 100), it is clearly seen that there is no existence of thick/ heavy nanoparticles bound with the $MoS_2$ monolayers, distinctly different with that in few-layer MoS$_2$. While here, due to the residing of unavoidable oxygen-containing PMMA (frequently found in Figure4a-4l) on graphene membranes used during the graphene transferring, it becomes intrinsically difficult for us to identify residual oxygen within the lattice of MoS$_2$ if any, particularly at the atomic level.

We then start to address the evolution of shape morphology and edge structure of MoS$_2$ monolayers, another key issue for the growth mechanism. Figure 4 displays a gallery of ADF-STEM images summarizing the size-dependent structure and morphology of monolayer MoS$_2$ samples grown on different regions of graphene. The smallest one found in our experiments with a diameter of close to 3.0 nm (Figure 4a), and an irregular polygon in shape. As the size increases, MoS$_2$ atomic clusters evolves into irregular hexagon shape (Figure 4f-h); and then into near-triangle shape with truncated corners (Figure 4i), before they finally form well-known triangular shape which are decorated mainly with Mo terminated zigzag edges[8-10, 30] (written as Mo-zz edge hereafter, see Figure 4k). Terraces and kinks (arrowed) can be frequently found along the edge of all MoS$_2$ monolayers, serving as chemically active sites for the epitaxy addition of Mo-S molecules or atomic clusters from the supply, either by direct deposition in gas phase or in solids phase after a surface diffusion process. From the results shown above, we can infer that the morphology of MoS$_2$ monolayers changes from irregular polygonal shape to triangular one with an increasing size (Figures 4a to 4j).

Accompanying the evolution of structure and mophology during the growth, the edge structure and termination also changed as shown in Figure 5. Here three MoS$_2$ monolayer sample with different characteristic morphology(irregular polygon, near-hexagon and near-triangle) were chosen as an example for ease of display (Figures 5a to 5c) and associated structural models (Figured 5d-5f). Over two primary types of edge structure were found: Mo-zz edge (highlighted with blue lines) and a bare Mo atom bound with zigzag terminated sulfur atoms, similar to the so-called Klein edge in graphite[31], written as S-Mo (marked by red lines in Figures 5a-5c) with the corresponding structural models shown as inset in Figure 5a (blue sites are Mo atoms and red sites are S atoms). A closer look at the S-Mo edge found that the out-extended Mo atoms tend to approach closer with its nearest neighbor, and thus a dimerization behaviors was observing on the S-Mo edge (inset in Figure 5a). For the smallest $MoS_2$ cluster (Figure 5a and 4b), these two types of edge structure are found approxiamtely equal number of quantities, and then the Mo-zz edge increase in its relative ratio as the size of $MoS_2$, and finally became predominant. Such an evolution can be more clealy read out from the quantitative analysis, the histograms shown in Figures 5g.

In order to explain the experimentally observed edge structures and their evolution, we further performed DFT simulations to calculate the formation energies of four zigzag edge structures (S-Mo, Mo-zz, S-zz (S terminated zigzag edges) and Mo-S (Mo terminated zigzag edges attached with two bare S atoms)) as a function of the chemical potential difference-$\Delta\mu_{Mo}$, $\Delta\mu_{Mo} = \mu_{Mo} - \mu_{Mo}(bulk)$), where $\mu_{Mo}$ and $\mu_{Mo}(bulk)$ are representing the chemical potential of Mo atom in the source precursor and in $MoS_2$ monolayer, respectively. Please refer to the computation details section in Supplementary Information for the details of modelling used in this study. The calculated results are shown in Figure 5h, from which one can read:(i) Mo-terminated edges (S-Mo and Mo-zz) have lower formation energies than those of S-terminated edges (S-S and Mo-S) under the Mo-rich condition (taken $\Delta\mu_{Mo}$> -0.3 eV), (ii) the formation energies of Mo-zz and S-Mo are quite close (i.e., differ only 0.01 eV at $\Delta\mu_{Mo}$= 0.1 eV).

In the next section, we discuss the microscopic growth process of $MoS_2$ monolayers, which seems to be different with that in few-layer case. Since no residual $MoO_xS_{2-y}$ nanostructures were observed on as-prepared $MoS_2$ monolayers as shown in Figure 1c and Figure 4, we can infer that the sublimated $MoO_2$ precursor should mostly form tiny molecular clusters ($MoO_2$, $(MoO_3)_n$ etc.), [32] and then sulfurized completely, followed by its deposition on graphene substrate as the nucleation center, or its addition onto as-formed nucleation center either from vapor ambient or via surface diffusion on graphene substrate. Relative ratios of these clusters can change as the carrier gas flow changes, the presence of $MoO_xS_{2-y}$ is reduced as the sulfur concentration in the reaction zone. As such, it can also well explain the difference in the as-formed products where few-layer $MoS_2$ or monolayer $MoS_2$ dominates, as shown in Figures 1b and 1c. According to the DFT calculations, suggesting that S-Mo and Mo-zz edges are dominant and with comparable ratio at the initial stage during the growth, as shown in Figure 4. Comparing to the Mo-zz edge, the S-Mo edge should possess relatively higher chemical reactivity due to the bare Mo atoms that will facilitate the incorporation of sourcing clusters, thus leading to a faster growth rate. According to the classic crystal growth theory, [33] fast growing edges/facets gradually disappear, while slow-growing edges/facets remain. In this case, it will eventually lead to the formation of triangle-shape MoS$_2$ monolayer decorated with Mo-zz edges, as most frequently observed. The slightly truncated shape observed on MoS$_2$ mono-layers shown in Figures 4j and 4k may be formed due to insufficient growth time, which again provides clear evidence for the proposed kinetics in edge structures, as discussed previously.

Figures 6a-6c present a summarized schematic diagram showing the microscopic process during the growth of few-layer MoS$_2$ under low gas flow(200 sccm) and Figures 6d-6f is the schematic of another route for the nucleation and growth kinetics of MoS$_2$ monolayers. And in our work, we did not focus on the transport process and there are two suggestions, possible loss of MoO$_x$S$_{2-y}$ nanoparticles during the process or gas deposition process. Li et al., proposed a three-step reaction pathway via examining the distribution of intermediate products in different forms. [34] From the results and discussions mentioned above, we could put forward some perspectives that may help us get high quality MoS$_2$. (1) A steady stream containing invariable partial pressure of molybdenum and sulfur containing vapor can also help us get a large area MoS$_2$ and at a desired relative ratio is essential towards the precisely controlled growth of MoS$_2$ atomic layers with desired edge structures and shape morphology. (2) For the precisely controlled layer thickness, extra care should be taken to control the sublimation of Mo precursor, and restrain the formation of larger MoO$_x$S$_{2-y}$ nanoparticles on the substrate for growth. (3) Given these considerations, the widely used strategy of the sulfurization of Mo-O precursor in a CVD furnace seems to be too simplified to fulfill this requirement. Further improvements including the use of metal-organic CVD (MOCVD) [35,36] or molecular beam epitaxy(MBE) [37] may be a better

solution in future.

In summary, we directly probe the nucleation seeds, evolution of edge structures and shape of evolution of pristinely-prepared MoS$_2$ materials. Two different types of nucleation centers were resolved: Mo-oxysulfide(MoO$_x$S$_{2-y}$) nanoparticles on few-layer samples, and pure molybdenum sulfide clusters for monolayer samples. Nevertheless, it was proposed for a growing MoS$_2$ monolayer that it grew from irregular polygonal-shaped cluster decorated with of S-Mo and Mo-zz edges in comparable ratio under the Mo-rich condition, then to triangle-like shapes with dominant Mo-zz edges with its size increasing. And this method would further provide us on deepened understandings of the growth mechanisms of atomically thin MoS$_2$ material via CVD, and other related two-dimensional TMD materials, and will pave the way for specific function and property-oriented growth under precisely control.

**Method**

Graphene films used here were grown on polycrystalline copper foils, and then transferred onto molybdenum (Mo) based TEM grids via a PMMA-assisted wet-chemistry process.[8] Graphene supported TEM grids were mounted onto a home-built ceramic carrier and loaded into a CVD system, facing down above an aluminum boat containing 1 mg of MoO$_2$ precursors (Sigma-Aldrich, 99%). Here we choose MoO$_2$, rather than MoO$_3$ as the precursor for two considerations because of the single-step chemical reaction MoO$_2$ +3S → MoS$_2$ + SO$_2$: 1) to reduce the reaction complexity, as a multi-step reduction reaction occurs for the sulfurization of MoO$_3$, 2) to grow high quality MoS$_2$ monolayer.[13] The whole CVD setup is shown in Figure 1a. Within a typical CVD process, the furnace was firstly heated to 300°C in 10 mins and hold for additional 10 mins, and then heated to 750°C in 40 mins and kept for next 25 mins. At about 15 mins after the furnace temperature reached 750°C, heating of 300mg of sulfur source (Aladdin, 99.999%) started with its temperature reaching 180°C in 2 mins, and then hold for next 10 mins. During the whole process, argon (99.999%) was used as the carrier gas, with an optimized flow rate of 200 standard-state cubic centimeter per minute (sccm) for growing few-layer samples and of 500 sccm for monolayer samples. The total growth time lasts for about 10 mins, and the furnace cooled down naturally. ADF-STEM was conducted with a FEI Chemi-STEM Titan G$^2$ 80-200 which was equipped with a probe-side spherical aberration-corrector and operated at an acceleration voltage of 200 kV. The convergent angle for illumination was set to 24 mrads with a probe current of 50-70 pA, and the collection angle was 50-100 mrads. Energy dispersive X-ray spectroscopy (EDS) was carried out on a Bruker super-X detection system.

**Acknowledgements**

We thank Prof. Feng Ding for his critical comments, and the Center of Electron Microscopy of Zhejiang University for the access to the microscope facilities. J.Y. acknowledges supports from Pao Yu-Kong International Foundation for a Chair Professorship in ZJU. H.B.S. acknowledged computational resources from the Shanghai Supercomputer Center.

**Competing Interests**

The authors declare no competing financial interests.

**Contributions**

C.J. conceived the project; D.Z. carried out most of the experiments (CVD, Raman and TEM), and analyzed the data with the assistance of F.J. V.A. and D.L., under the supervision of C.J.; H.S. carried out the DFT calculations; O.O. J. Y. and Z.Z. contributed to the discussions; D.Z. and C.J. co-wrote the paper with the inputs and suggestions from others.

**Funding**

This work was financially supported by the National Basic Research Program of China (Grant No. 2014CB932500 and No. 2015CB921004) and the National Science Foundation of China (Grant No. 51472215, No. 51222202, and No. 11404309).

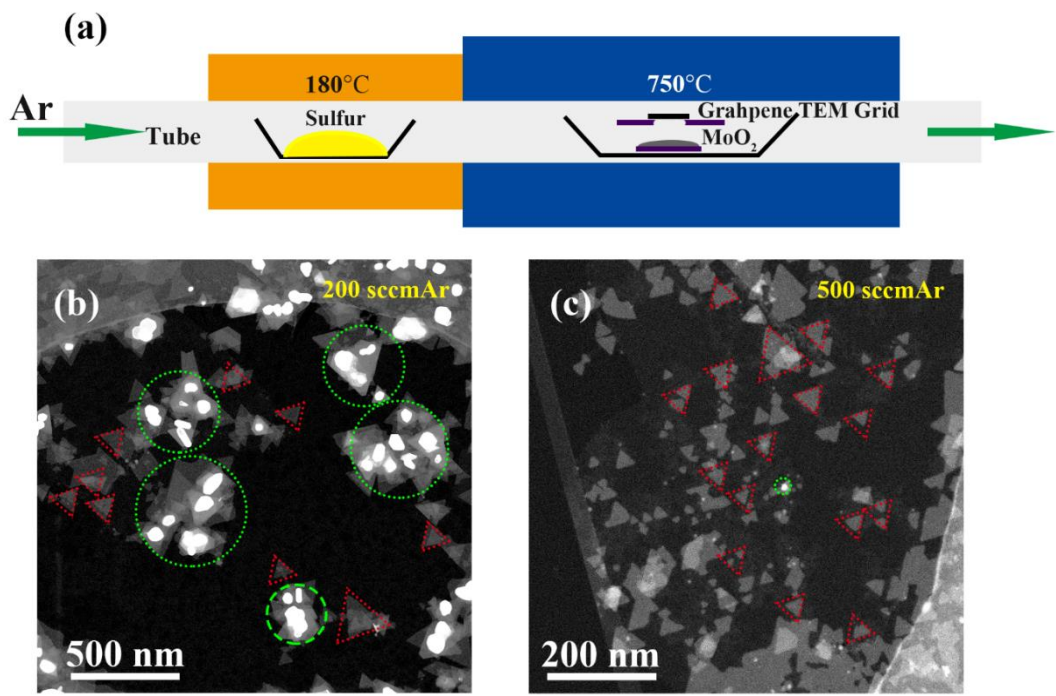

**Figure 1.** (a) Schematic illustration of our CVD system in which graphene supported TEM grids were placed face down to the MoO$_2$-containing boat. (b) A typical low-magnification ADF-STEM image representing as-grown products on graphene supports at 200sccm Ar. The green dotted circles indicate few-layer MoS$_2$ bound with thick/heavy nanoparticle (and nanorods), while the red dotted circles show monolayer MoS$_2$. (c) A typical low-magnification ADF-STEM images representing as-grown products at 500 sccm Ar, where monolayer MoS$_2$ predominates (in red dotted circles).

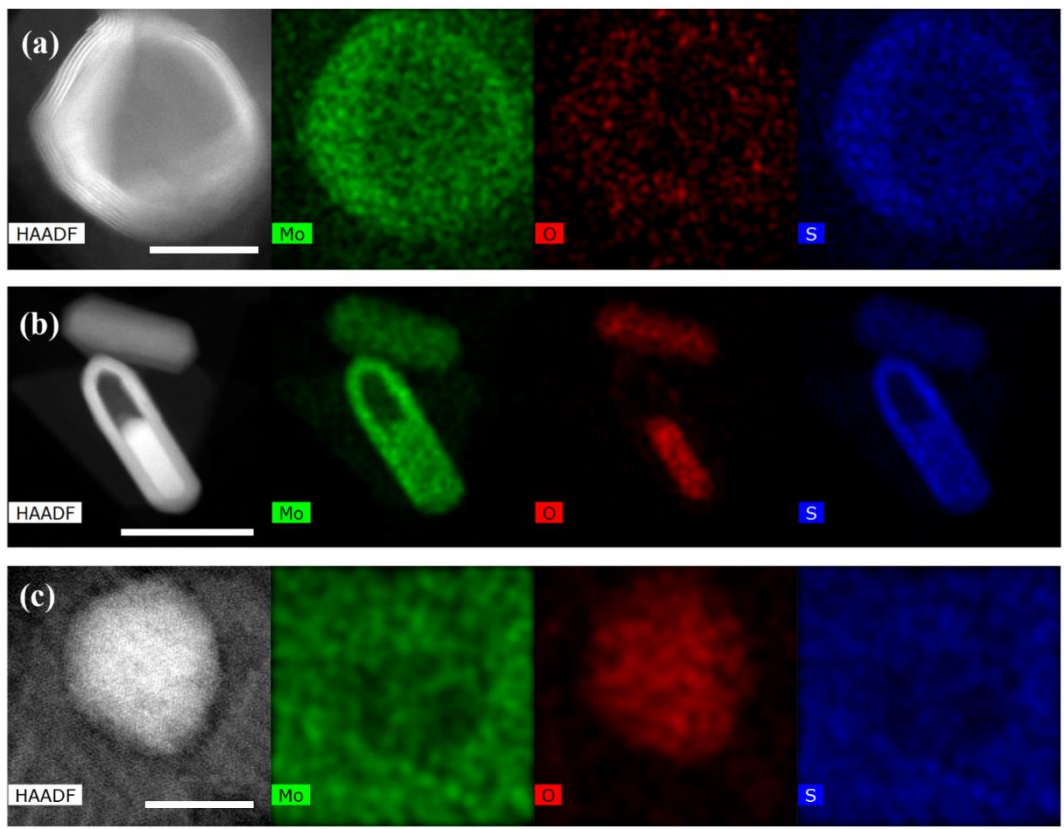

**Figure 2.** STEM-EDS mapping showing the atomic structural and chemical composition of thick structures on few-layer MoS2: (a) a multi-shelled fullerene with nearly empty core, (b) a tubular like structure with partly scrificed core, (c) a core without fullerene structure (scale bar: 20 nm)

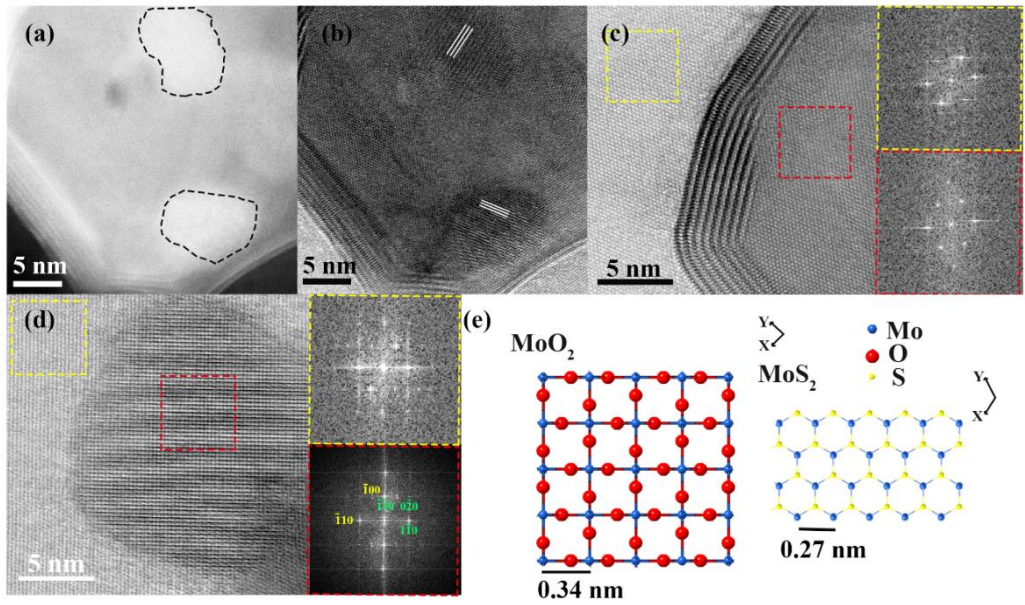

**Figure 3.** (a) ADF-STEM image of typical nanoparticles with the core-shell fullerene structure. (b) The corresponding BF-STEM image of typical nanoparticles with the core-shell fullerene structure. (c) BF-STEM image of structure of empty area of the core. (d) BF-STEM image of nanoparticles without the core-shell fullerene structure and the corresponding FFT patterns collected from selected regions on few-layer $MoS_2$ (yellow dotted circle) and MoOS nanoparticle (red dotted circle). (e) Sschematic represents the formation of $MoS_2$ from $MoO_2$ in a CVD process.

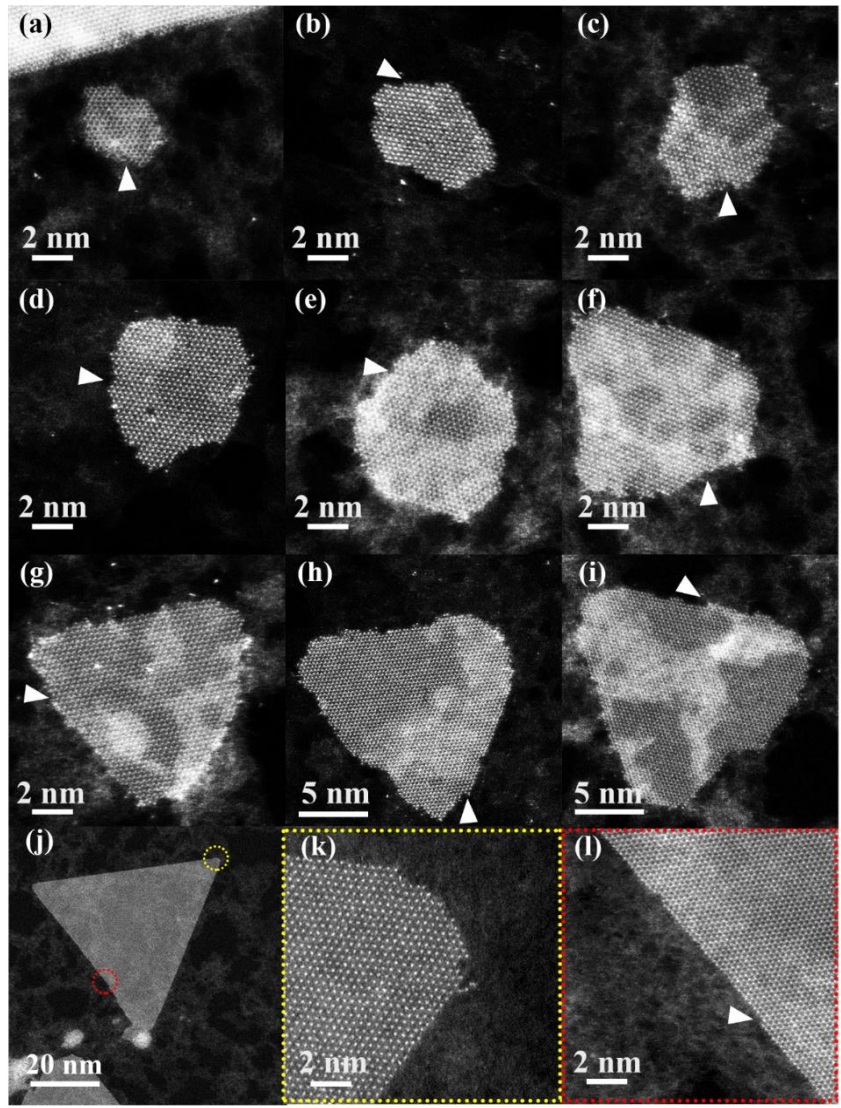

**Figure 4.** (a-i) ADF-STEM images of monolayer $MoS_2$ with different sizes found on graphene, illuminating the shape evolution during CVD growth. A number of terraces with atomic steps can be found on the edge (arrowed, not all arrowed). (j) Low-mag ADF-STEM image of a triangle-shaped $MoS_2$ monolayer, (k) atomic structure of the region highlighted in yellow dotted circle, (l) atomic structure of the local region highlighted in red dotted circle.

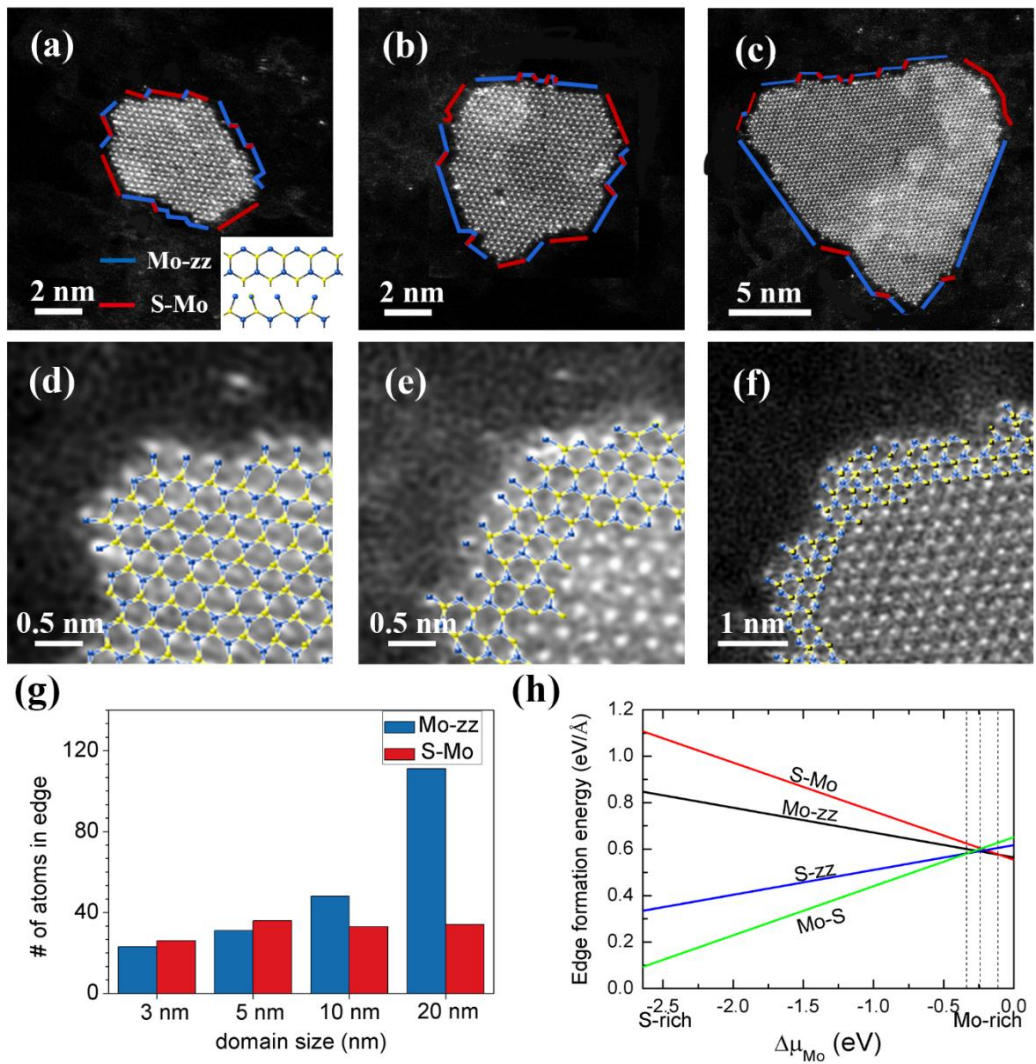

**Figure 5.** (a) ADF-STEM image of irregular shaped monolayer $MoS_2$ cluster with two primary types of edge structure: Mo-zz edge and S-Mo edge (denoted by blue and red lines separately). (b) ADF-STEM image of hexagonal-shaped $MoS_2$ (c) ADF-STEM image of triangle $MoS_2$. (d) The edge structural model of the irregular $MoS_2$, (e)the edge structural model of the near-hexagon $MoS_2$. (f) the edge structural model of the near-triangle $MoS_2$. (g) Histogram of the amount of the edge atoms with different size. (h) Results from DFT calculations for the formation energies of different edges as a function of $\Delta\mu_{Mo}$.

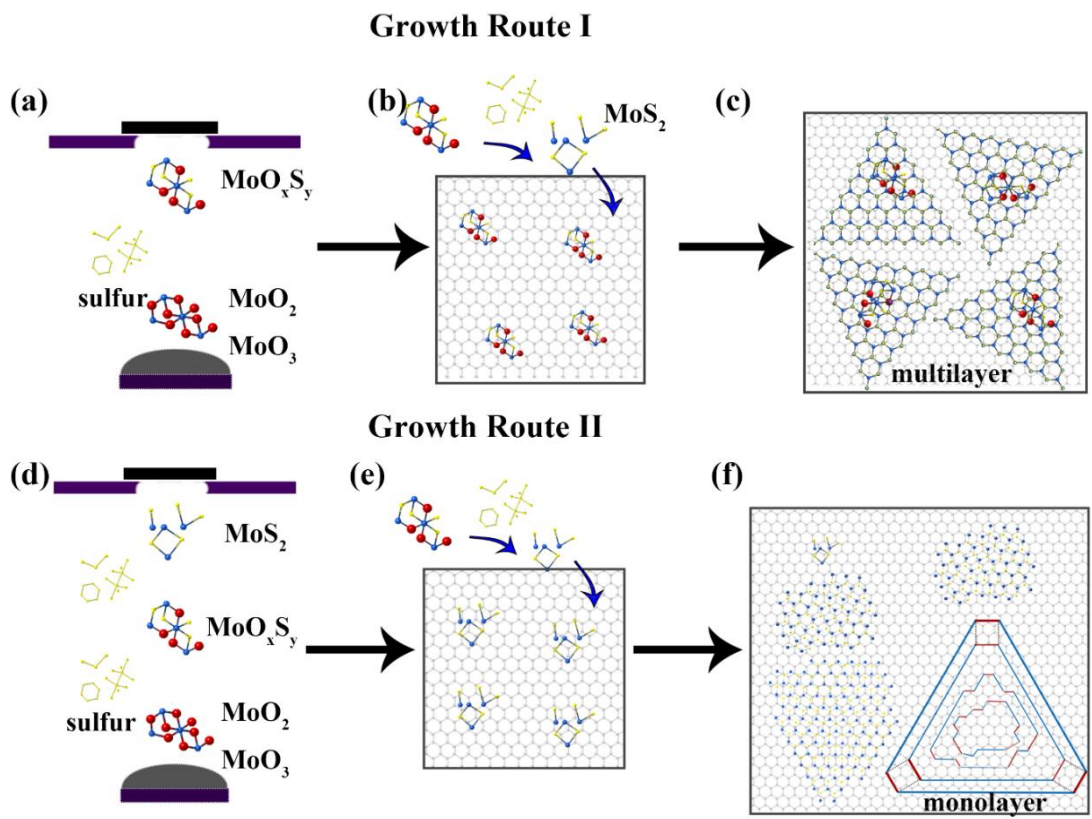

**Figure 6.** (a-c)Schematic sketches explaining the possible route for the nucleation and growth kinetics of few-lay MoS₂ bound with thick core, and that (d-f) for the nucleation and growth kinetics of MoS₂ monolayers.